\documentstyle[12pt]{article}

\begin{document}
\centerline{\bf Phase of Aharonov-Bohm oscillations in
conductance of mesoscopic systems}
\centerline{P. S. Deo$^1$  and A. M. Jayannavar$^1$}
\centerline{International Center for Theoretical Physics,
Trieste, Italy}
\vskip .2in

Motivated by a recent experiment we analyze in detail the phase
of Aharonov-Bohm oscillations across a 1D ring with a stub
coupled to one of its arms, in the presence of a magnetic flux.
We demonstrate that there are two kinds of conductance extremas.
One class of them are fixed at particular flux values and can
only change abruptly from a maxima to a minima as incident
energy is varied. We show a different mechanism for such abrupt
phase change in conductance oscillation.  We demonstrate that
these extremas can exhibit $``$phase locking".  However, the
second kind of extremas can shift continuously as the incident
energy is varied.

\vfill
\eject

Transport in mesoscopic samples has been studied extensively
over the last decade. Technological developments has helped us
to fabricate samples of sizes smaller than the single particle
coherence length of the electron.  Quantum mechanical
interference effects drastically affect the transport in such
systems and many non trivial phenomenon has been observed.  Of
particular interest are the universal conductance fluctuations,
quantization of point contact conductance \cite{book1}, normal
state Aharonov-Bohm effect, \cite{ww} current magnification
effect, \cite{cuma} etc,.  A recent experiment \cite{ya} reports
some striking features of the conductance across an
Aharonov-Bohm (A-B) ring with a quantum dot situated in one of
its arms.  The observations of the experiment are as follows.
Transport through the dot in the Coulomb blockade regime has a
coherent component.  The phase of conductance oscillations
change by $\pi$ over a finite energy scale as the Fermi energy
or incident energy crosses the resonances of the dot.  This
scale is much smaller (an order of magnitude smaller) than the
scale in which the phase of the wavefunction changes as the
Fermi energy crosses the resonance of the dot.  The observations
were theoretically analyzed in ref \cite{ye,wi,nl}.  It has been
argued \cite{ye} that if at all the phase of conductance
oscillations change, the change should be absolutely sharp i.e.,
should take place over an energy scale of zero width, or else it
would mean a break down of micro-reversibility.  However a
finite width was indeed observed in the experiment. In ref.
\cite{ye} this was accounted to effects like noise and
fluctuations. Other works \cite{nl} attribute the finite width
to the non-linear response.  Ref \cite{ye} tried to demonstrate
such an abrupt change in the phase of conductance oscillations
or parity of conductance oscillations \cite{ye} by a model
calculation in 1D.  The effect can be physically explained as
follows.  A-B effect of normal electrons in a ring cannot be
interpreted in terms of partial waves propagating along the two
arms of the ring.  The ring has some bound states that exhibit
strong resonance phenomenon in transport when the ring is
coupled to leads. The resonances shift with the magnetic field.
If it shifts away from the Fermi energy then conductance will
decrease, and vice versa.  So there will be an abrupt change of
$\pi$ in phase of conductance oscillations as the Fermi energy
crosses the resonance of a ring.

We give a completely different mechanism of such an abrupt
change in the parity of conductance oscillations. This mechanism
exhibits the change in parity as the Fermi energy crosses the
resonance of the dot (as observed in the experiment) and not the
resonance of the ring.  We believe that this point is important
for the following three reasons.  i) Thermal smearing length in
the experiment was estimated to be comparable to the ring's
length and it is unlikely that the resonances of the ring can
manifest themselves in the real situation. However, the study of
these resonances \cite{ye} help us in understanding of the phase
changes in the conductance oscillations.  ii) If the phase of
conductance oscillations is determined by the bound states of
the isolated system then we do not expect any regular behavior
at consecutive resonances because the E versus $\alpha$ curve
of two consecutive bound states need not have opposite slopes
\cite{ram}. iii) Our mechanism is related to the coherent
scattering by the actual geometry of the dot used in the
experiment. So, it is likely that such parity changes are
present in the actual experiment.  iv) Besides we show a
possible cause of not observing abrupt change of parity of
conductance oscillations within the framework of Landauer
formula without the violation of microreversibility.

A schematic diagram of the system on which we do the model
calculation is shown in fig. 1. A one dimensional ring is
connected to two ideal leads on two sides. The other ends of the
ideal leads are connected to a reservoir.  If the chemical
potential of the reservoir on two sides are unequal then there
will be a transport current through the ring. A stub or a side
arm is situated on one arm of the ring. A flux $\phi$ pierce
through the center of the ring.  As a result an electron picks
up a phase $\alpha=2\pi \phi/\phi_0$ in going round the ring
once. Here $\phi_0=hc/e$ is the elementary flux quantum.  The
total circumference of the ring is $l=p+q+r$, where the various
lengths p, q and r are denoted in the figure. The length of the
stub is L. Throughout in our calculations we have chosen the
parameters $p/l=.25$, $q/l=.25$, $r/l=.5$ and $L/l$=1. The
geometry is a one dimensional representation of the actual
system with which the experiment was performed. Persistent
currents in such coupled geometries has been already studied in
some details \cite{cou}. We solve for the transmission across
this system exactly using the free electron wave guide theory on
networks \cite{ne}. We use Griffth's boundary conditions at the
junction of the stub and the ring and the hard wall boundary
condition at the end of the stub.  This allow us to calculate
the scattering matrix at the junctions from the first
principles.  Total transmission can be calculated analytically,
Expression being too long to reproduce here, we analyze our
results graphically. Transmission coefficient is directly
related to the transmission conductance G by Landauer formula
and dimensionless conductance is $g=G/(2e^2/h) T$.

In fig 2 we have plotted the conductance (g) versus $\alpha$ for
three values of incident energy or dimensionless Fermi wave
vector kl i.e., kl=($\pi-.01)$ (thin line), $\pi$ (thick line),
and $(\pi+.01)$ (thickest line). At kl=$\pi$ the isolated stub
has a bound state. We find that at the exact value kl=$\pi$
there is no conductance oscillation. On the two sides of
kl=$\pi$ i.e., at kl=$\pi+.01$ and kl=$\pi-.01$ the conductance
oscillations are out of phase by $\pi$ or belong to opposite
parity class.  Both have a periodicity of $2\pi$ and both are
symmetric in $\alpha$ as required by microreversibility. This
change of parity occurs for infinitesimal change of Fermi energy
kl on the two sides of the value kL=$\pi$ i.e., the change of
parity occurs over an energy scale of zero width. Note that at
$\alpha$=0 ring has no bound state at kL=$\pi$ but the nearest
ones are at kl=0 and $2\pi$. This sudden phase change can be
understood if we map the stub into a delta function potential.
It is known that the delta function potential is V(x)=k Cot(kL)
$\delta(x-x_0)$ \cite{deo} where $x_0$ is the position of the
stub. At kL=$\pi$ the effective potential is infinite. This
implies that there is no propagation along the dot arm of the
ring and so no interference induced by the magnetic flux. This
is why the conductance oscillations disappear for this
particular value of the Fermi energy. On the lower side of this
value i.e., at kl=$\pi-.01$, the effective potential is an
attractive delta function potential and on the higher side of
this value of the Fermi energy i.e., at kl=$\pi+.01$ it is a
repulsive delta function potential. This discontinuous change in
the strength of the effective potential changes the phase of the
wavefunction on the upper arm discontinuously by $\pi$ and hence
changes the parity of the conductance oscillations. Note that
this sudden phase change also breaks the parity effect of
persistent currents in isolated system \cite{deo}. But this
parity change in conductance oscillations is not related to the
slope of the eigenenergy.

In a certain Fermi energy range before the resonance is crossed
the energy dependent effective delta function potential changes
very rapidly because cot(kL) change very rapidly around the
value kl=$n\pi$. This also happens in a certain energy range
after the resonance is crossed. This rapid change in the
effective delta function potential incorporates large phase
changes in the wave function of the electron. The phase of
transmission amplitude across the stub is $\theta=
arctan(cot(kL)/2)$ and it is plotted in fig. 3 in the range
kl=$\pi+.01$ to kl=$\pi+.13$ i.e., on the higher side of the
resonance at kl=$\pi$.  The figure shows that the phase changes
by a large amount in this range but the change is less than
$\pi$. Surprisingly, this phase change which is less than $\pi$,
cannot change the phase of the conductance oscillations.  This
is shown in fig. 4. Both for kl=$\pi+.01$ (thin line) and
kl=$\pi+.13$ (thick line) the maximas occur in the same
positions and so do the minimas.  The phase of the conductance
oscillations get locked and cannot be affected by the phase
changes created by the stub. This $``$phase locking" was
conjectured in ref \cite{ye} and we can demonstrate it with an
explicit example.  Due to the non-locality of the electron,
change in effective potential in the dot arm not only changes
the phase of the wavefunction in the dot arm but also the phase
of the electron wavefunction in the other arm.  These phase
changes balance each other in such a way that the parity of the
conductance oscillations remain unchanged. Thus at these
extremas the parity of the conductance oscillations can only
change abruptly independent of the amount of phase acquired in
traversing the dot. If this were true at all maximas and all
minimas then it is not possible to see a finite width in which
the parity of the conductance oscillations change. But in the
following we show that there can be some extremas that do not
exhibit $``$phase locking" and can shift their positions
gradually as the incident energy is varied, without violating
the condition that the conductance is symmetric in flux.

However if we increase the Fermi energy further away from $\pi$
then additional conductance extremas can appear. This is shown
in fig 5 where we have plotted the conductance (g) versus
$\alpha$ for kl=$\pi-.2$ (thin line) and kl=$\pi+.2$ (thick
line). We find that for kl=$\pi+.2$ at the point marked A there
is a maxima at which there is no extrema for kl=$\pi-.2$.  So
there can be new maximas opposite to which there is no minima
and vice-versa. These new extremas are not restricted by
micro-reversibility and can change in any fashion. They can
change continuously in contrast to the extremas considered in
figure 2 that can only change abruptly.  This is shown in fig. 6
where we plot conductance (g) versus $\alpha$ for kl=$\pi+.2$
(thin line) and kl=$\pi+.3$ (thick line). The maximas are
slightly shifted with respect to each other. Sometimes these new
extremas are very prominent compared to the extremas that remain
fixed in position.  When the thermal smearing width is as large
as the separation between resonances the small resonances may
merge with the more prominent ones. If the resonances that
survive are the ones that can shift their positions as the
incident energy is varied, then it is not unlikely that maximas
change to a minima in a finite range of incident energy.

At this point we want to stress that we are not interested in
the actual origin of these second kind of extremas.  There can
be various reasons for their origin.  Depending on the Fermi
energy sometimes the conductance oscillations can show a
$\phi_0/2$ periodicity and sometimes a $\phi_0$ periodicity
\cite{past}. So if for some energies the conductance oscillates
more with the flux than at some other Fermi energy then we can
have maximas opposite to which there is not always a minima and
vice versa.  These additional extremas can also appear because
as the energy is varied the strength of the effective delta
potential also varies which results in changing the effective
arm lengths of the ring. So, the system itself changes
effectively as the incident energy changes. However, the point
that we want to stress is that microreversibility does not
guarantee that the extremas are always fixed in positions. Had
they been so it would not be possible to see the parity of
conductance oscillations changing over a finite energy range
within the frame work of Landauer conductance formula.

Finally in figure 7 we plot the conductance (g) versus incident
wave vector kl when the flux through the ring is zero. It shows
a very complicated sequence of conductance maximas and also
conductance zeros. But they do not match with the resonances of
the ring or zeros of the stub. The leads and the coupling of the
dot to the ring changes them drastically.  For example the
isolated dot exhibits zero conductance at kl=$\pi$ but the
combined system show a large transmission.
 
The main result of our work can be summarized as follows. The
conductance oscillations can exhibit two kinds of extremas.  One
class of them can change their parity abruptly as the incident
energy is varied or cannot change their parity at all.  The
other class of extremas can change their parity continuously.
Our study makes it possible in principle to observe the effect
as observed in the experiment of Yacoby et al. We also show a
different mechanism for abrupt change in parity of conductance
oscillations.

One of us (PSD) thanks R. Parwani for useful discussions.

\vfill
\eject
{\bf Figure Captions}

Fig. 1. Open metallic ring coupled to two electron reservoirs in
the presence of magnetic flux $\phi$. A side arm of length L is
connected to one of the arms.

Fig. 2. Plot of conductance versus $\alpha$ for various values
of incident dimensionless wave vector kl.

Fig. 3. Plot of the phase of the transmission amplitude $\theta$
across the isolated stub versus kl.

Fig. 4. Plot of conductance versus $\alpha$

Fig. 5. Plot of conductance versus $\alpha$

Fig. 6. Plot of conductance versus $\alpha$

Fig. 7. Plot of conductance versus kl in the absence of magnetic
flux.

\begin{thebibliography}{99}
\bibitem{pa} permanent address:
Institute of Physics, Bhubaneswar 751005, India.
\bibitem{book1} {\it Quantum Coherence in Mesoscopic Systems}, Vol
254 of NATO advanced study Institute, Series B; Physics,
edited by Kramer B (Plenum, New York) 1991.
\bibitem{ww} Washburn S and Webb. R. A, Adv. Phys. {\bf
35} 375(1986).
\bibitem{cuma} A. M. Jayannavar and P. Singha Deo, Phys. Rev. B
{\bf 51}, 10175(1995); T. P. Pareek, P. Singha Deo and A. M.
Jayannavar, Phys. Rev. B, {\bf 52} 14657(1995).
\bibitem{ya} A. Yacoby, M. Heiblum, D. Mahalu and H. Shtrikman
Phys. Rev. Lett. {\bf 74} 4047(1995).
\bibitem{ye} L. Yeyati and M. B$\ddot u$ttiker Phys. Rev. B
{\bf 52} R14360(1995)
\bibitem{wi} G. Hackenbroich and H. A. Weidenm$\ddot u$ller
(unpublished)
\bibitem{nl} C. Bruder, R. Fazio and H. Schoeller (unpublished).
\bibitem{ram} P. A. Sreeram and P. Singha Deo (unpublished).
\bibitem{cou} P. Singha Deo, Phys. Rev B {\bf 52} 5441(1995);
T. P. Pareek and A. M. Jayannavar, Phys. Rev. B (in press);
F. Pasceud and Montambaux (unpublished).
\bibitem{ne} A. M. Jayannavar and P. Singha Deo, Mod. Phys.
Lett B, {\bf 8} 301(1994);
A. M. Jayannavar and P. Singha Deo, Phys Rev. B
{\bf 49} 13685(1994).
P. Singha Deo and A. M. Jayannavar, Phys. Rev. B {\bf 50},
11629(1994)
\bibitem{deo} P. Singha Deo Phys. Rev. B (in press)
\bibitem{past} J. D'Amato, H. M. Pastawski and J. F. Weisz,
Phys. Rev. B {\bf 39}, 3554(1989).

\end{thebibliography}
\end{document}